\documentclass[english,aps,prper,reprint,showpacs,titlepage,longbibliography]{revtex4-2}

\usepackage[T1]{fontenc}	
\usepackage[latin9]{inputenc}	
\usepackage{geometry}		
\usepackage{graphicx}
\usepackage[above,below]{placeins}	
\usepackage{times}
\usepackage{amsmath}
\usepackage{array}

\usepackage{float}

\usepackage{verbatim}
\usepackage{hyperref}  
\hypersetup{colorlinks=true,urlcolor=blue,citecolor=blue,linkcolor=blue}   
\usepackage{enumitem}          
\setlist{nosep}                 

\begin{document}

\begin{titlepage}

\title{Calculation of the transient response of lossless transmission lines}



\author{Takamitsu Koyano}
\author{Jake S.\ Bobowski}	
\affiliation{Department of Physics, University of British Columbia, 3333 University Way, Kelowna, British Columbia, V1V 1V7,
    Canada} 


  \begin{abstract}
  We present an analytical calculation of the transient response of ideal (i.e.\ lossless) transmission lines.  The calculation presented considers a length of transmission line connected to a signal generator with output impedance $Z_\mathrm{g}$ and terminated with a load impedance $Z_\mathrm{L}$.  The approach taken is to analyze a circuit model of the system in the complex-frequency or $s$-domain and then apply an inverse Laplace transform to recover the time-domain response.  We consider both rectangular pulses and voltage steps (i.e.\ the Heaviside function) applied to the input of the transmission line.  Initially, we assume that $Z_\mathrm{g}$ and $Z_\mathrm{L}$ are purely real/resistive. At the end of the paper, we demonstrate how the calculations can be generalized to consider reactive impedances.    
  \end{abstract}

 \maketitle
\end{titlepage}
\section{Introduction}

Despite their apparent simplicity, transmission lines are rich in physics, can be challenging to analyze, and can lead to complex behaviors~\cite{Steer:2019, Bobowski:2021}.  Here, we will focus on developing analytical solutions for the transient response of ideal or lossless transmission lines.  One approach is to use so-called ``bounce'' diagrams in which the reflections at the two boundaries of the transmission line are analyzed~\cite{Haus:1989}.  A second approach is to analyze the transmission line in the frequency domain and then use an inverse Fourier transform to deduce the time-domain response~\cite{Aitken:1970}.  Below, we present a variation of the this latter approach using inverse Laplace transforms and include some of the details that have been omitted from Ref.~\onlinecite{Aitken:1970}.  

\section{Transmission Lines}\label{sec:trans}

\begin{figure}[htb]
\centering
\includegraphics[width=\linewidth]{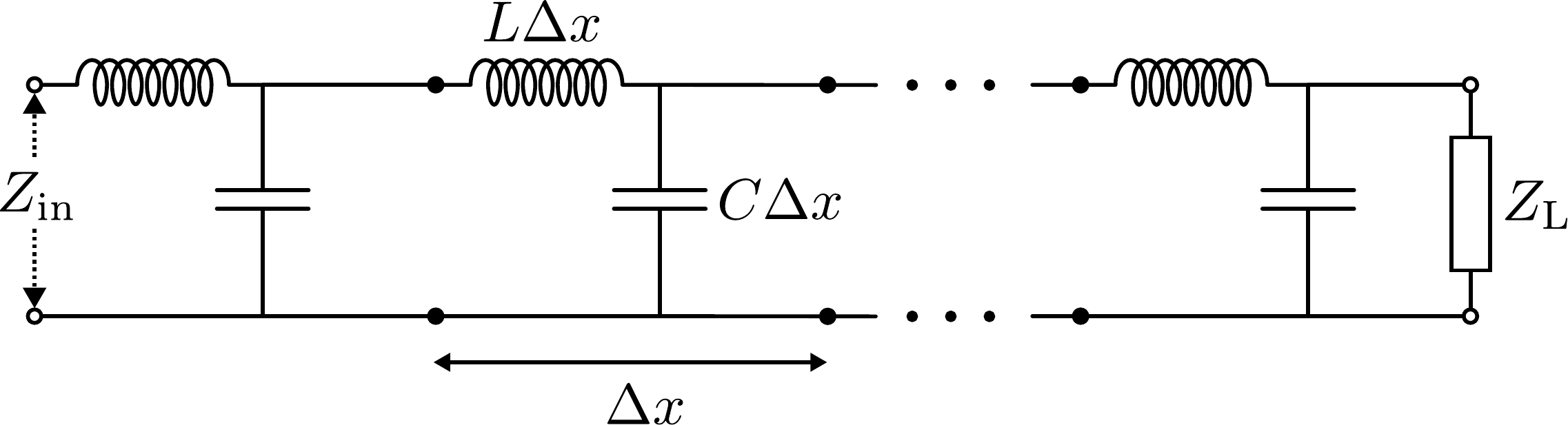}
  \caption{Lumped-element circuit model of an ideal transmission line terminated with load impedance $Z_\mathrm{L}$.\label{fig:lossless}}
\end{figure}

This section presents the lumped-element circuit model of an ideal transmission line and summarizes some of the important results that follow from an analysis of the model.  Transmission line effects become important when the wavelength $\lambda$ of a signal propagating along a transmission is less than or comparable to its length $\ell$.  In this case, the voltage and current varies along the length of the transmission line and, in order to develop an lumped-element circuit model, it is first necessary to divide the line into $N$ identical sections of length $\Delta x$ such that $\lambda\gg\Delta x$.  As shown in Fig.~\ref{fig:lossless}, each section of the transmission line is then model as a series inductance $L\Delta x$ and a shunt capacitance $C\Delta x$.  Here, $L$ and $C$ represent the per-unit-length inductance and capacitance of the transmission line, respectively~\cite{Steer:2019, Haus:1989}.   

An analysis of one section of transmission line model in the limit $\Delta x\to 0$ leads to the following pair of coupled partial differential equations for the voltage $v(x,t)$ and current $i(x,t)$:
\begin{align}
\frac{\partial v(x,t)}{\partial x} &= -L\frac{\partial i(x,t)}{\partial t},\\
\frac{\partial i(x,t)}{\partial x} &= -C\frac{\partial v(x,t)}{\partial t}.
\end{align}
If harmonic time dependencies \mbox{$v(x,t)=V(x)e^{j\omega t}$} and \mbox{$i(x,t) = I(x)e^{j\omega t}$} are assumed, the pair of first-order partial differential equations decouple and can be expressed as independent second-order ordinary differential equations, one for the voltage amplitude $V(x)$ and the other for the current amplitude $I(x)$: 
\begin{align}
\frac{d^2 V(x)}{d x^2} &= -\beta^2 V(x),\label{eq:V}\\
\frac{d^2 I(x)}{d x^2} &= -\beta^2 I(x),\label{eq:I}
\end{align}
where $\beta=\omega\sqrt{LC}$ is the propagation constant and \mbox{$v_0=\sqrt{LC}$} is the signal propagation speed.  Equations (\ref{eq:V}) and (\ref{eq:I}) are the familiar wave equations and have solutions:
\begin{align}
V(x) &= V_+e^{-j\beta x} + V_-e^{j\beta x},\\
I(x) &= \frac{1}{Z_\mathrm{c}}\left[V_+e^{-j\beta x} - V_-e^{j\beta x}\right].\label{eq:current}
\end{align}
Here, $Z_\mathrm{c}=\sqrt{L/C}$ is known as the characteristic impedance of the transmission line and $V_+$ and $V_-$ are the coefficients of the forwards and backwards traveling components of $v(x,t)$ and $i(x,t)$, respectively.  Notice that the transmission line shown in Fig.~\ref{fig:lossless} is terminated by a load impedance $Z_\mathrm{L}$.  When a forward-traveling signal is incident on $Z_\mathrm{L}$, in general, it will be partially absorbed by $Z_\mathrm{L}$ and partially reflected with the reflection coefficient given by:
\begin{equation}
\Gamma_\mathrm{L} = \frac{Z_\mathrm{L} - Z_\mathrm{c}}{Z_\mathrm{L} + Z_\mathrm{c}}.\label{eq:Gamma}
\end{equation}  
When $Z_\mathrm{L} = Z_\mathrm{c}$, the load impedance is matched to the characteristic impedance and there is no reflection.  On the other hand, when $Z_\mathrm{L} = \infty$ (open circuit), $\Gamma_\mathrm{L} = 1$ which corresponds to a perfect reflection.  Likewise, $Z_\mathrm{L} = 0$ (short circuit) gives $\Gamma_\mathrm{L} = -1$ which also corresponds to a perfect reflection.  In this case, however, the reflected signal is inverted relative to the incident signal~\cite{Steer:2019, Haus:1989}. 

We conclude this section by recalling one final important result from an analysis of the circuit model of Fig.~\ref{fig:lossless}.  The equivalent impedance $Z_\mathrm{in}$ looking into a lossless transmission line of length $\ell$ terminated by load impedance $Z_\mathrm{L}$ is given by:
\begin{equation}
Z_\mathrm{in}=Z_\mathrm{c}\frac{Z_\mathrm{L} + jZ_\mathrm{c}\tan\left(\beta\ell\right)}{Z_\mathrm{c} + jZ_\mathrm{L}\tan\left(\beta\ell\right)}.
\end{equation}
In the sections that follow, we will be analyzing transmission line circuits in the complex frequency domain \mbox{$s = j\omega$}.  $Z_\mathrm{in}$ can be re-expressed in terms of $s$ by recognizing that \mbox{$\beta = -js/v_0$} and $\tan\left(\beta\ell\right) = -j\tanh\left(s\ell/v_0\right)$ such that~\cite{Steer:2019, Haus:1989}:
\begin{equation}
Z_\mathrm{in}=Z_\mathrm{c}\frac{Z_\mathrm{L} + Z_\mathrm{c}\tanh\left(s\ell/v_0\right)}{Z_\mathrm{c} + Z_\mathrm{L}\tanh\left(s\ell/v_0\right)}.\label{eq:Zin}
\end{equation}

\section{Laplace Transforms}\label{sec:Laplace}
As shown by Eq.~(\ref{eq:Zin}), a length of transmission line terminated by load impedance $Z_\mathrm{L}$ can be expressed as a single impedance $Z_\mathrm{in}$ in the $s$-domain.  When solving for the transient response of a transmission line, we will first analyze an equivalent problem in the $s$-domain and then use an inverse Laplace transform to find the desired solution in the time domain.  A Laplace transform takes a function of time $f(t)$ and converts it a function of $s$: $F(s) = \mathcal{L}\{f(t)\}$.  The mathematical definition of the Laplace transform is:
\begin{equation} 
F(s) = \int_0^\infty e^{-st}f(t)dt.\label{eq:laplace}
\end{equation}
Likewise, the inverse Laplace transform converts a function of the complex frequency $F(s)$ to a function a time: \mbox{$f(t) = \mathcal{L}^{-1}\{F(s)\}$}~\cite{Schiff:1900}.

In Sec.~\ref{sec:transient}, we will make use of a translation theorem for inverse Laplace transform.  In order to derive this theorem, we will first prove a corresponding theorem for the Laplace transform.  Namely, the goal is to show that:
\begin{equation}
\mathcal{L}\{u(t - t_0)f(t-t_0)\} = e^{-st_0}F(s),\label{eq:translation}
\end{equation}
where:
\begin{equation}
u(t - t_0) =
    \begin{cases}
      1 & t > t_0\\
      0 & t < t_0,
    \end{cases}    
\end{equation}
is the Heaviside function.  To prove Eq.~(\ref{eq:translation}), we use the definition of the Laplace transform:
\begin{align}
\mathcal{L}\{u(t - t_0)f(t-t_0)\} &= \int_0^\infty e^{-st}u(t - t_0)f(t - t_0)dt\\
&= \int_{t_0}^\infty e^{-st}f(t - t_0)dt,
\end{align}
where the second equality is due to the fact that $u(t - t_0) = 0$ over the time range $0 < t < t_0$.  To finish the proof, a change of variables $\tau = t - t_0$ is made which results in:
\begin{align}
\mathcal{L}\{u(t - t_0)f(t-t_0)\} &= \int_0^\infty e^{-s(\tau + t_0)}f(\tau)d\tau\\
&= e^{-st_0}\int_0^\infty e^{-s\tau}f(\tau)d\tau\\
&= e^{-st_0}F(s),\label{eq:trans1}
\end{align}
where, in the last step, we have used the definition of the Laplace transform given by Eq.~(\ref{eq:laplace}).  Finally, taking the inverse Laplace transform of Eq.~(\ref{eq:trans1}) gives:
\begin{equation}
\mathcal{L}^{-1}\{e^{-st_0}F(s)\}=u(t - t_0)f(t - t_0)\label{eq:invTrans}
\end{equation}
which is the desired translation theorem for the inverse Laplace transform~\cite{Schiff:1900}.

\section{Transmission Line analysis in the Complex-Frequency Domain}\label{sec:s-domain}
\begin{figure}[htb]
\centering
\includegraphics[width=\linewidth]{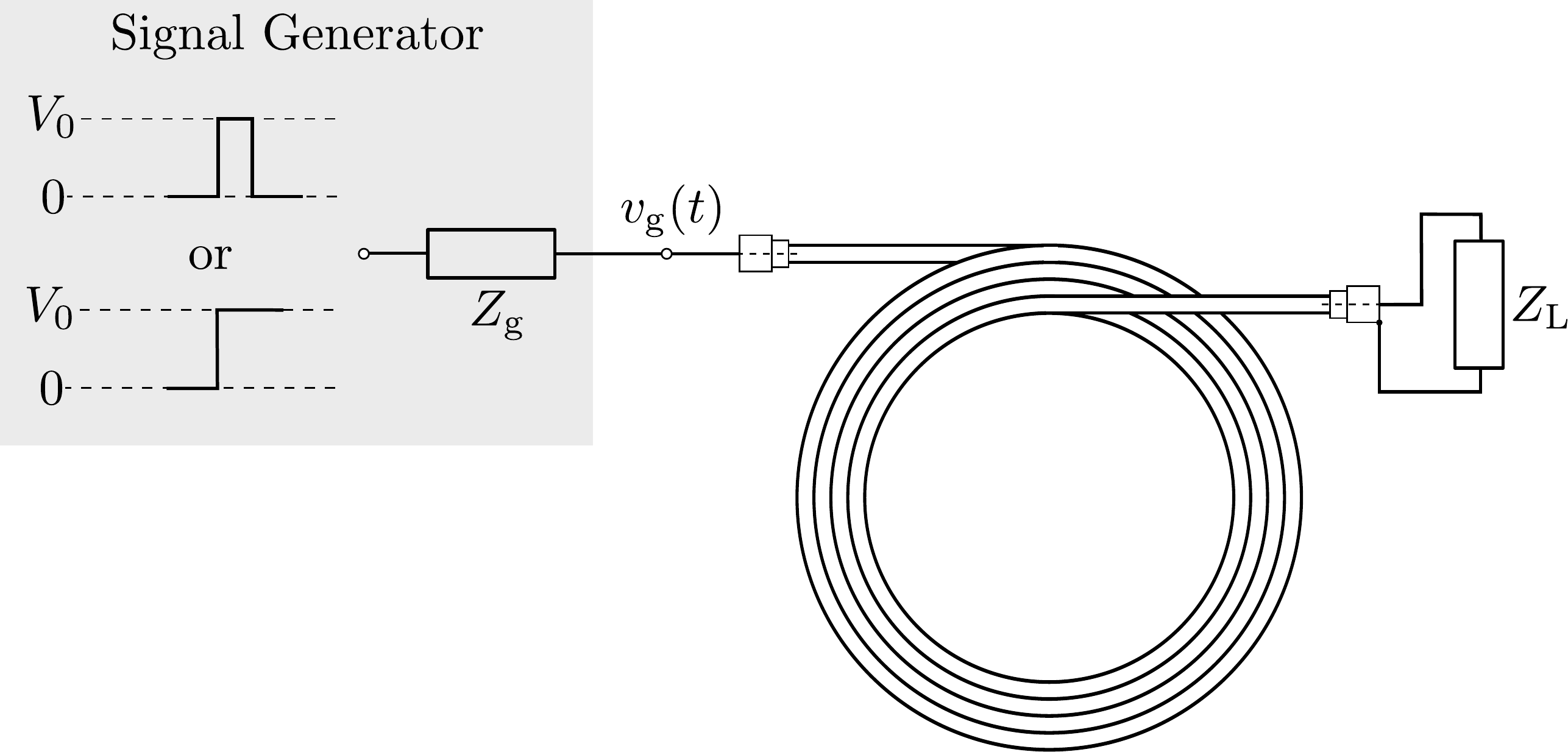}
  \caption{The transmission line geometry to be analyzed.  A length of coaxial cable connected to a signal generator with output impedance $Z_\mathrm{g}$ and terminated by load impedance $Z_\mathrm{L}$.\label{fig:coax}}
\end{figure}

Figure~\ref{fig:coax} shows the geometry of the problem under consideration.
A signal generator with output impedance $Z_\mathrm{g}$ supplies a voltage (typically, a rectangular pulse or a step) at one end of a length of transmission line while the opposite end is terminated by a load impedance $Z_\mathrm{L}$.  The goal is to calculate $v_\mathrm{g}(t)$ in the time domain at the junction between the signal generator and the input of the transmission line.

\begin{figure}[htb]
\centering
\includegraphics[width=0.45\linewidth]{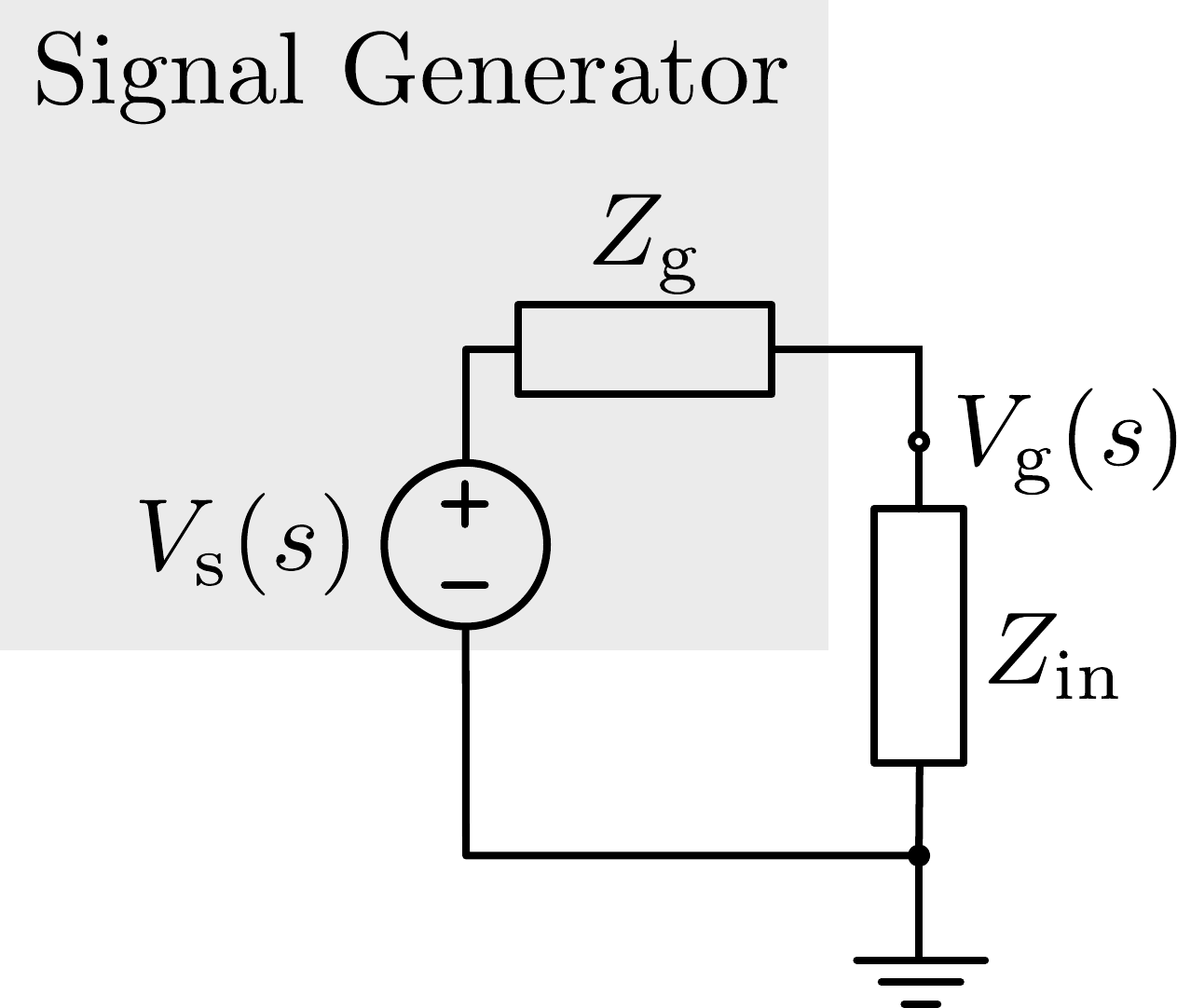}
  \caption{The equivalent $s$-domain circuit of Fig.~\ref{fig:coax}.  $Z_\mathrm{in}$ is given by Eq.~(\ref{eq:Zin}) and $V_\mathrm{s}(s)$ is the $s$-domain representation of the signal generator output.\label{fig:equiv}}
\end{figure}

An equivalent circuit representation of Fig.~\ref{fig:coax} in the $s$-domain is shown in Fig.~\ref{fig:equiv}.
In this figure, $V_\mathrm{s}(s) = \mathcal{L}\{v_\mathrm{s}(t)\}$ is the $s$-domain representation of the signal generator output.  Table~\ref{tab:vs} gives the time and $s$-domain representations of a rectangular pulse and a voltage step.
\begin{table}
\begin{tabular}{m{1.8cm} m{3.3cm} m{0.23cm} m{2.65cm}}
~ & time domain & ~ & $s$-domain\\
\hline\hline\\[-2ex]
\raisebox{-.35\height}{\includegraphics[width=1.8 cm]{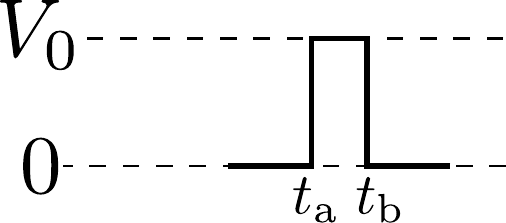}} & $V_0\left[u(t - t_\mathrm{a}) - u(t - t_\mathrm{b})\right]$ & $\Leftrightarrow$ & $\dfrac{V_0}{s}\left(e^{-st_\mathrm{a}} - e^{-st_\mathrm{b}}\right)$\\
~\\[-1ex]
\raisebox{-.35\height}{\includegraphics[width=1.8 cm]{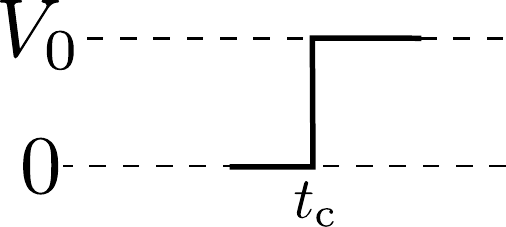}} & \qquad\qquad\qquad $V_0 u(t - t_\mathrm{c})$ & $\Leftrightarrow$ & $\dfrac{V_0 e^{-st_\mathrm{c}}}{s}$
\end{tabular}
\caption{The time and $s$-domain representations of the rectangular pulse and voltage step.  Both are assumed to have an amplitude of $V_0$.  The right-most column depicts the shape of the signal in the time domain.\label{tab:vs}}
\end{table}
Both the pulse and step are assumed to have an amplitude $V_0$.  The pulse starts at time $t=t_\mathrm{a}$ and ends at time $t=t_\mathrm{b}$.  For the step, the change in voltage occurs at time $t=t_\mathrm{c}$.  The equivalent circuit of Fig.~\ref{fig:equiv} is a simple voltage divider such that:
\begin{equation}
V_\mathrm{g}(s) = V_\mathrm{s}(s)\frac{Z_\mathrm{in}/Z_\mathrm{g}}{1+ Z_\mathrm{in}/Z_\mathrm{g}}.
\end{equation}
Subbing in Eq.~(\ref{eq:Zin}) for $Z_\mathrm{in}$ and multiplying through by the denominator of $Z_\mathrm{in}$ leads to:
\begin{equation}
\frac{V_\mathrm{g}(s)}{V_\mathrm{s}(s)} = \frac{\dfrac{Z_\mathrm{L}}{Z_\mathrm{g}} + \dfrac{Z_\mathrm{c}}{Z_\mathrm{g}}\tanh\left(\dfrac{s\ell}{v_0}\right)}{\left(1 + \dfrac{Z_\mathrm{L}}{Z_\mathrm{g}}\right) + \left(\dfrac{Z_\mathrm{L}}{Z_\mathrm{c}} + \dfrac{Z_\mathrm{c}}{Z_\mathrm{g}}\right)\tanh\left(\dfrac{s\ell}{v_0}\right)}
\end{equation}
Expressing $\tanh(s\ell/v_0)$ in terms of exponentials, multiplying through by the denominator of $\tanh(s\ell/v_0)$, and collecting like terms results in:
\begin{equation}
\frac{V_\mathrm{g}(s)}{V_\mathrm{s}(s)} = \frac{\left(\dfrac{Z_\mathrm{L} + Z_\mathrm{c}}{Z_\mathrm{g}}\right) + \left(\dfrac{Z_\mathrm{L} - Z_\mathrm{c}}{Z_\mathrm{g}}\right)e^{-2s\ell/v_0}}{q_+ + q_- e^{-2s\ell/v_0}},\label{eq:VgVs2}
\end{equation}
where:
\begin{align}
q_\pm &\equiv 1 \pm \dfrac{Z_\mathrm{L}}{Z_\mathrm{c}} + \dfrac{Z_\mathrm{L} \pm Z_\mathrm{c}}{Z_\mathrm{g}}\\
& = \frac{\left(Z_\mathrm{c} \pm Z_\mathrm{g}\right)\left(Z_\mathrm{L} \pm Z_\mathrm{c}\right)}{Z_\mathrm{c}Z_\mathrm{g}}.\label{eq:q_pm}
\end{align}
Subbing Eq.~(\ref{eq:q_pm}) in Eq.~(\ref{eq:VgVs2}) and dividing through by $\left(Z_\mathrm{c} + Z_\mathrm{g}\right)\left(Z_\mathrm{L} + Z_\mathrm{c}\right)$ gives:
\begin{equation}
\frac{V_\mathrm{g}(s)}{V_\mathrm{s}(s)} = \frac{\dfrac{Z_\mathrm{c}}{Z_\mathrm{g} + Z_\mathrm{c}}\left[1 + \left(\dfrac{Z_\mathrm{L} - Z_\mathrm{c}}{Z_\mathrm{L} + Z_\mathrm{c}}\right)e^{-2s\ell/v_0}\right]}{1 - \left(\dfrac{Z_\mathrm{g} - Z_\mathrm{c}}{Z_\mathrm{g} + Z_\mathrm{c}}\right)\left(\dfrac{Z_\mathrm{L} - Z_\mathrm{c}}{Z_\mathrm{L} + Z_\mathrm{c}}\right)e^{-2s\ell/v_0}}.
\end{equation}
Identifying $\Gamma_\mathrm{L}$ from Eq.~(\ref{eq:Gamma}) and defining:
\begin{equation}
\Gamma_\mathrm{g} = \frac{Z_\mathrm{g} - Z_\mathrm{c}}{Z_\mathrm{g} + Z_\mathrm{c}},
\end{equation} 
which accounts for reflections of backwards traveling signals at the transmission line input, allows one to write:
\begin{equation}
\frac{V_\mathrm{g}(s)}{V_\mathrm{s}(s)} = \frac{\dfrac{Z_\mathrm{c}}{Z_\mathrm{g} + Z_\mathrm{c}}\left(1 + \Gamma_\mathrm{L}e^{-2s\ell/v_0}\right)}{1 - \Gamma_\mathrm{g}\Gamma_\mathrm{L}e^{-2s\ell/v_0}}.\label{eq::VgVs3}
\end{equation}

Since $\Gamma_\mathrm{g}\Gamma_\mathrm{L}e^{-2s\ell/v_0}$ is necessarily less than one, the denominator of Eq.~(\ref{eq::VgVs3}) can be expressed as a geometric series:
\begin{equation}
\left(1 - \Gamma_\mathrm{g}\Gamma_\mathrm{L}e^{-2s\ell/v_0}\right)^{-1} = \sum_{n = 0}^\infty\left(\Gamma_\mathrm{g}\Gamma_\mathrm{L}e^{-2s\ell/v_0}\right)^n.
\end{equation}
As a result, $s$-domain representation of the desired signal can be expressed as:
\begin{multline}
V_\mathrm{g}(s) = V_\mathrm{s}(s)\left(\frac{Z_\mathrm{c}}{Z_\mathrm{g} + Z_\mathrm{c}}\right)\left(1 + \Gamma_\mathrm{L}e^{-2s\ell/v_0}\right) \\ \cdot\sum_{n = 0}^\infty\left(\Gamma_\mathrm{g}\Gamma_\mathrm{L}e^{-2s\ell/v_0}\right)^n.\label{eq:VgVs4}
\end{multline} 
Finally, writing out the first several terms of the sum reveals that $V_\mathrm{g}(s)$ can be re-expressed as:
\begin{multline}
V_\mathrm{g}(s) = V_\mathrm{s}(s)\left(\frac{Z_\mathrm{c}}{Z_\mathrm{g} + Z_\mathrm{c}}\right) \\ \cdot \left[1+ \left(1 + \Gamma_\mathrm{g}\right)\sum_{k = 0}^\infty \Gamma_\mathrm{g}^k\Gamma_\mathrm{L}^{k+1}e^{-2\left(k+1\right)s\ell/v_0}\right].\label{eq:VgVs5}
\end{multline}

\section{Transient Response}\label{sec:transient}
In this section, inverse Laplace transforms of $V_\mathrm{g}(s)$ are used to find $v_\mathrm{g}(t)$ in the time domain.  We will first consider the case in which $Z_\mathrm{g}$ and $Z_\mathrm{L}$ are purely resistive, and hence $\Gamma_\mathrm{g}$ and $\Gamma_\mathrm{L}$ are independent of $s$.  At the end of this section, we will consider an example in which the load impedance is reactive which will lead to a reflection coefficient $\Gamma_\mathrm{L}$ that depends on the complex frequency $s$.

\subsection{Resistive $Z_\mathrm{g}$ and $Z_\mathrm{L}$}\label{sec:resistive}
Taking the inverse Laplace transform of Eq.~(\ref{eq:VgVs5}) and factoring quantities that are independent of $s$ out of the transform yields:
\begin{multline}
v_\mathrm{g}(t) = \left(\frac{Z_\mathrm{c}}{Z_\mathrm{g} + Z_\mathrm{c}}\right)\Bigg[\mathcal{L}^{-1}\left\{V_\mathrm{s}(s)\right\} + \left(1 + \Gamma_\mathrm{g}\right)\Bigg. \\
\Bigg. \cdot\sum_{k = 0}^\infty \Gamma_\mathrm{g}^k\Gamma_\mathrm{L}^{k+1}\mathcal{L}^{-1}\left\{e^{-2\left(k+1\right)s\ell/v_0}V_\mathrm{s}(s)\right\}\Bigg].
\end{multline}
$\mathcal{L}^{-1}\{V_\mathrm{s}(s)\}$ is simply $v_\mathrm{s}(t)$ and, from the translation theorem given by Eq.~(\ref{eq:invTrans}):
\begin{multline}
\mathcal{L}^{-1}\left\{e^{-2\left(k+1\right)s\ell/v_0}V_\mathrm{s}(s)\right\}\\
= u\left[t - 2\left(k+1\right)\ell/v_0\right]v_\mathrm{s}\left[t - 2\left(k+1\right)\ell/v_0\right].
\end{multline}
Since both $u(t)$ and $v_\mathrm{s}(t)$ are zero for $t<0$, the result above can simply be expressed as $v_\mathrm{s}\left[t - 2\left(k+1\right)\ell/v_0\right]$, which is just a copy of the original pulse delayed by a time \mbox{$\tau_k = 2\left(k+1\right)\ell/v_0$}.  Therefore, the general solution for the transient response of a transmission line for the case of resistive impedances $Z_\mathrm{g}$ and $Z_\mathrm{L}$ is:
\begin{multline}
v_\mathrm{g}(t) = \left(\frac{Z_\mathrm{c}}{Z_\mathrm{g} + Z_\mathrm{c}}\right)\bigg\{v_\mathrm{s}(t) + \left(1 + \Gamma_\mathrm{g}\right)\bigg. \\
\bigg. \cdot\sum_{k = 0}^\infty \Gamma_\mathrm{g}^k\Gamma_\mathrm{L}^{k+1}v_\mathrm{s}\left[t - 2\left(k+1\right)\ell/v_0\right]\bigg\}.\label{eq:solnA}
\end{multline}
A nice feature of this solution is that truncating the sum at \mbox{$k = N$} only limits the range of times over which the solution is valid.  The solution $v_\mathrm{g}(t)$ remains exact for times \mbox{$t < t_\mathrm{max} = t_0 + 2\left(N + 1\right)\ell/v_0$}, where $t_0 = t_\mathrm{a}$ for a pulse input and $t_0 = t_\mathrm{c}$ for a step input.  The only effect of including more terms in the sum is to extend the value of $t_\mathrm{max}$.

\begin{figure*}[htb]
\centering
\begin{tabular}{ccc}
(a) \includegraphics[height=0.85\columnwidth]{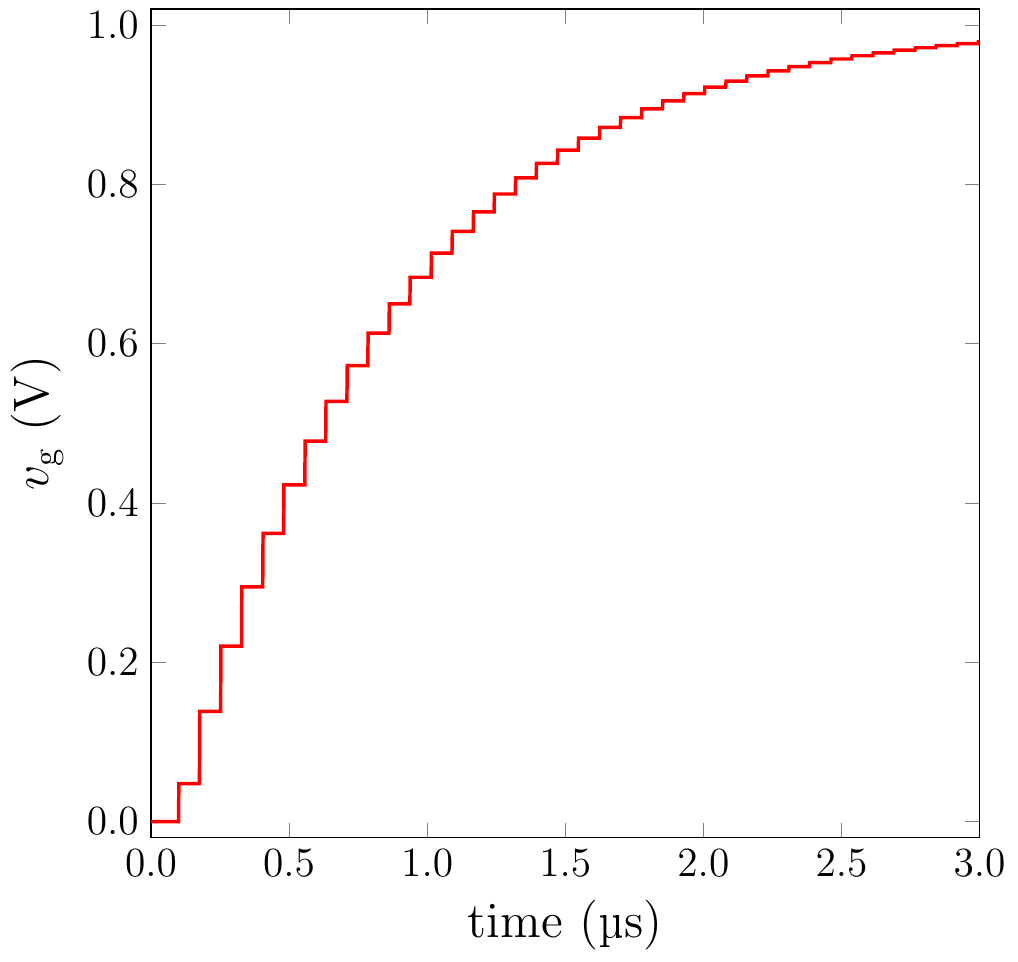} & ~\qquad~ & (b) \includegraphics[height=0.85\columnwidth]{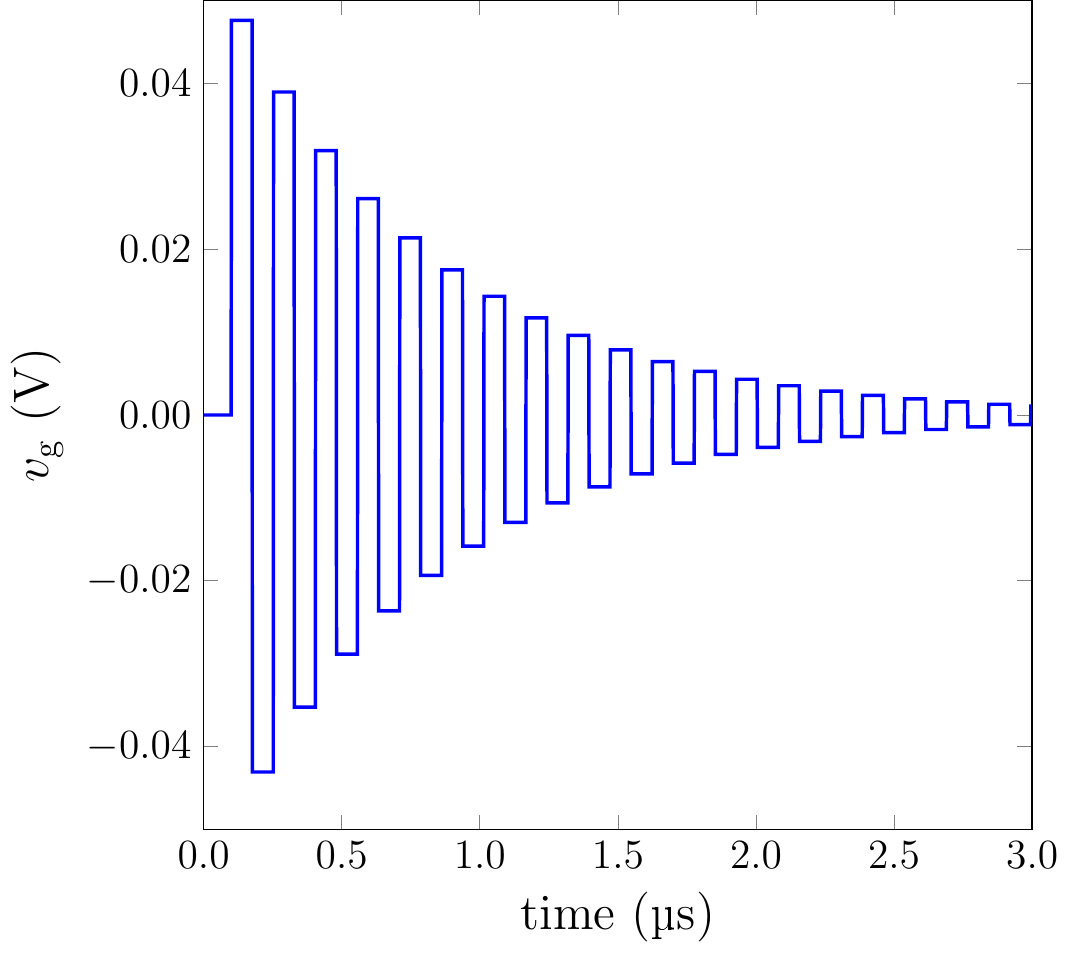}
\end{tabular}
\caption{Transient response of a transmission line to a voltage step with $Z_\mathrm{g}=1\rm\ k\Omega$ and (a) an open-circuit and (b) a short-circuit termination.\label{fig:step}}
\end{figure*}

To a demonstration a few a well-known cases, we will assume parameter values of $\ell = 8\rm\ m$, $Z_\mathrm{c} = 50\rm\ \Omega$, and $v_0 =0.7c$, where $c$ is the vacuum speed of light. The first case considered is $Z_\mathrm{g}=1\rm\ k\Omega$ and $Z_\mathrm{L}\to\infty$ (open circuit).  At the open end of the transmission line, forward-traveling signals experience a perfect reflection with $\Gamma_{L} = 1$ while backwards-traveling signals are partially reflected with $\Gamma_\mathrm{g} = 0.905$.  Subbing these values into Eq.~(\ref{eq:solnA}) and taking $v_\mathrm{s}(t)$ to be  a voltage step with $V_0 = 1\rm\ V$ and $t_\mathrm{c} = 0.1\rm\ \mu s$ results in the transient response shown in Fig.~\ref{fig:step}(a)~\cite{Haus:1989}.

The resulting ``staircase'' pattern is due to repeated reflections at both ends of the transmission line.  The time between steps is $2\ell/v_0$, the time required for a signal to travel the length of the transmission line and back again.  Figure \ref{fig:step}(a) shows that the height of the staircase steps diminish as time increases.  This effect is due to the fact that the signal is partially absorbed each time it is reflected at $Z_\mathrm{g}$ since $\left\vert \Gamma_\mathrm{g}\right\vert < 1$.  

In Fig.~\ref{fig:step}(b), we show that transient response to a voltage step for a transmission line with a short-circuit termination $Z_\mathrm{L} = 0$ (all other parameters are the same as used in part (a) of the figure).  The time between steps remains the same and the diminishing step size is clearly visible.  In this case, however, at each step, the transient signal reverses sign since $\Gamma_\mathrm{L} = -1$ which inverts the signal at each reflection from $Z_\mathrm{L}$.

\subsection{Matching the impedance $Z_\mathrm{g}$ to $Z_\mathrm{c}$}\label{sec:matched}
When the output impedance $Z_\mathrm{g}$ of the signal generator is matched to the characteristic impedance of the transmission line, $\Gamma_\mathrm{g} = 0$ and the form of $v_\mathrm{g}(t)$ given by Eq.~(\ref{eq:solnA}) simplifies considerably, since only the $k = 0$ term from the sum survives:
\begin{equation}
v_\mathrm{g}(t) = \frac{1}{2}\left[v_\mathrm{s}(t) + \Gamma_\mathrm{L}v_\mathrm{s}\left(t - 2\ell/v_0\right)\right].\label{eq:solnB}
\end{equation}
The first term in Eq.~(\ref{eq:solnB}) is the incident pulse and the second term is the reflected pulse which has been scaled by $\Gamma_\mathrm{L}$ and delayed by $2\ell/v_0$, the time required for the signal to travel twice the length of the transmission line.  For an open-circuit termination, $\Gamma_\mathrm{L} = 1$ and the reflected pulse is an exact copy of the incident pulse.  On the other hand, for a short-circuit termination, $\Gamma_\mathrm{L} = -1$ and the reflected pulse is inverted relative to the incident pulse.  

\begin{figure*}[htb]
\centering
\begin{tabular}{ccc}
(a) \includegraphics[height=0.85\columnwidth]{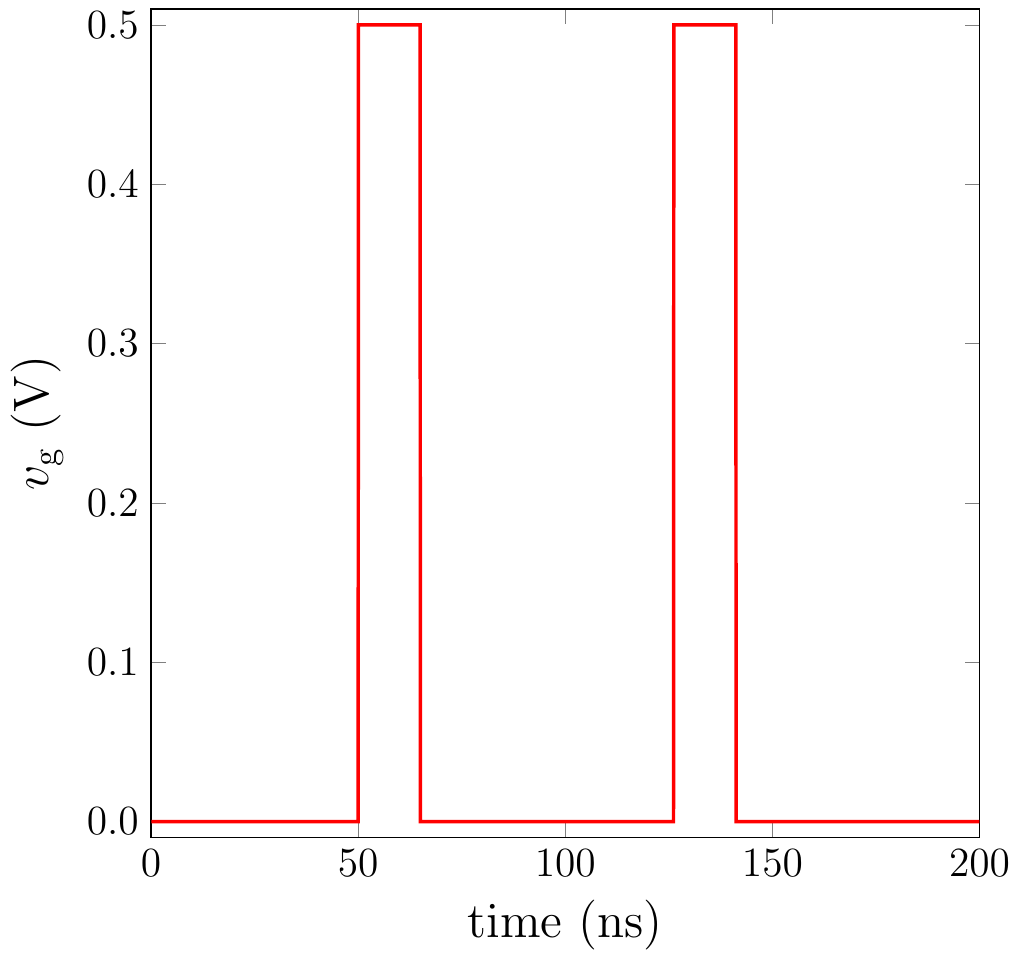} & ~\qquad~ & (b) \includegraphics[height=0.85\columnwidth]{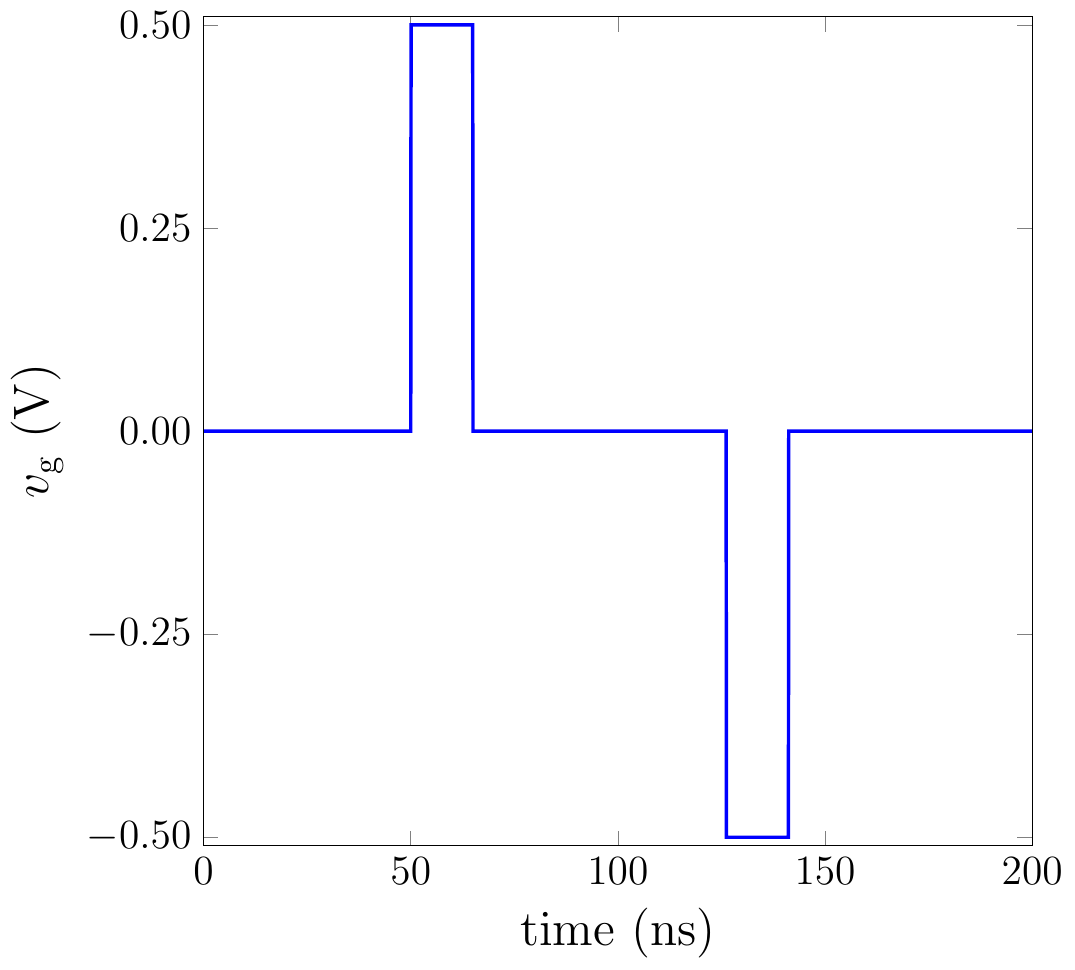}
\end{tabular}
\caption{Transient response of a transmission line to a voltage pulse with $Z_\mathrm{g}=50\rm\ \Omega$ and (a) an open-circuit and (b) a short-circuit termination.\label{fig:pulse}}
\end{figure*}

Figure~\ref{fig:pulse} shows the transient response for the case of (a) an open-circuit termination and (b) a short-circuit termination.
Other than the value of $Z_\mathrm{g}$ and the form of $v_\mathrm{s}(t)$, the parameters used to generate these plots were the same as those used in Sec.~\ref{sec:resistive}.  The output of the signal generator was taken to be a rectangular pulse starting at $t_\mathrm{a} = 50\rm\ ns$ with a height $V_0 = 1\rm\ V$ and a width of $15\rm\ ns$ ($t_\mathrm{b} = 65\rm\ ns$).  The output impedance of the signal generator was assumed to be $50\rm\ \Omega$, which is equal to the characteristic impedance of the transmission line.

For the open-circuit case ($\Gamma_\mathrm{L} = 1$), as shown by Eq.~(\ref{eq:solnB}), two identical pulse of height $V_0/2$ are produced.  The second pulse, which was reflected from the open circuit, is delayed by a time of $2\ell/v_0$ with respect to the initial incident pulse.  For the short circuit case, the only difference is that the reflected pulse is inverted due to the fact that, in this case, $\Gamma_\mathrm{L} = -1$.  

A pulse reflected from $Z_\mathrm{L} = 0$ must be inverted to satisfy the requirement that the voltage across the short circuit must be zero.  When the incident and reflected pulse overlap at the position of the short, they must sum to zero.  In a similar way, the current at an open circuit must be zero.  Equation~(\ref{eq:current}) shows that the total current at any time $t$ and position $x$ is determined by the difference between the forward and backward traveling signals.  Therefore, when the identical incident and reflected pulses overlap at the open circuit, the net current will be zero.

\subsection{Reactive load impedance $Z_\mathrm{L}$}\label{sec:reactive}
We now provide an example calculation of the transient response of a lossless transmission line when the load impedance $Z_\mathrm{L}$ depends on the complex frequency $s$.  In general, one needs to return to Eq.~(\ref{eq:VgVs5}) and insert the correct form of $\Gamma_\mathrm{L}(s)$ and, if necessary, $Z_\mathrm{g}(s)$ and $\Gamma_\mathrm{g}(s)$.  For this example calculation, like in Sec.~\ref{sec:matched}, we will assume that $Z_\mathrm{g} = Z_\mathrm{c}$ such that $\Gamma_\mathrm{g} = 0$ and $Z_\mathrm{c}/\left(Z_\mathrm{g} + Z_\mathrm{c}\right) = 1/2$.  In this case, Eq.~(\ref{eq:VgVs5}) becomes:
\begin{equation}
V_\mathrm{g}(s) = \frac{1}{2}V_\mathrm{s}(s)\left[1 + \Gamma_\mathrm{L}(s)e^{-2s\ell/v_0}\right].\label{eq:ZL0}
\end{equation}

We first consider the product $V_\mathrm{s}(s)\Gamma_\mathrm{L}(s)$ for an inductive load impedance and a voltage step of height $V_0$.  In this case, $V_\mathrm{s}(s) = V_0e^{-st_\mathrm{c}}/s$ and $Z_\mathrm{L}(s) = sL$ such that:
\begin{align}
V_\mathrm{s}(s)\Gamma_\mathrm{L}(s) &= \dfrac{V_0}{s}\left(\dfrac{sL - Z_\mathrm{c}}{sL + Z_\mathrm{c}}\right)e^{-st_\mathrm{c}}\\
&=\dfrac{V_0}{s}\left[\dfrac{sL}{sL + Z_\mathrm{c}} - \dfrac{Z_\mathrm{c}}{sL + Z_\mathrm{c}}\right]e^{-st_\mathrm{c}}\\
&= V_0\left[\dfrac{1}{s + \tau_\mathrm{L}^{-1}} - \frac{\tau_\mathrm{L}^{-1}}{s\left(s + \tau_\mathrm{L}^{-1}\right)}\right]e^{-st_\mathrm{c}},\label{eq:ZL1}
\end{align}
where the an inductive time constant $\tau_\mathrm{L} = L/Z_\mathrm{c}$ has been defined.  Using a partial fraction decomposition, the second term within the square brackets of Eq.~(\ref{eq:ZL1}) can be re-expressed as:
\begin{equation}
\frac{\tau_\mathrm{L}^{-1}}{s\left(s + \tau_\mathrm{L}^{-1}\right)} = \frac{1}{s} - \frac{1}{s + \tau_\mathrm{L}^{-1}},
\end{equation} 
such that:
\begin{equation}
V_\mathrm{s}(s)\Gamma_\mathrm{L}(s) = V_0\left[\dfrac{2}{s + \tau_\mathrm{L}^{-1}} - \frac{1}{s}\right]e^{-st_\mathrm{c}}.
\end{equation}
Substituting this result into Eq.~(\ref{eq:ZL0}) gives:
\begin{equation}
V_\mathrm{g}(s) = \frac{V_\mathrm{s}(s)}{2} + \frac{V_0}{2}\left[\dfrac{2}{s + \tau_\mathrm{L}^{-1}} - \frac{1}{s}\right]e^{-\left(t_\mathrm{c} + 2\ell/v_0\right)s}.\label{eq:ZL3}
\end{equation}

\begin{figure*}[htb]
\centering
\begin{tabular}{ccc}
(a) \includegraphics[height=0.9\columnwidth]{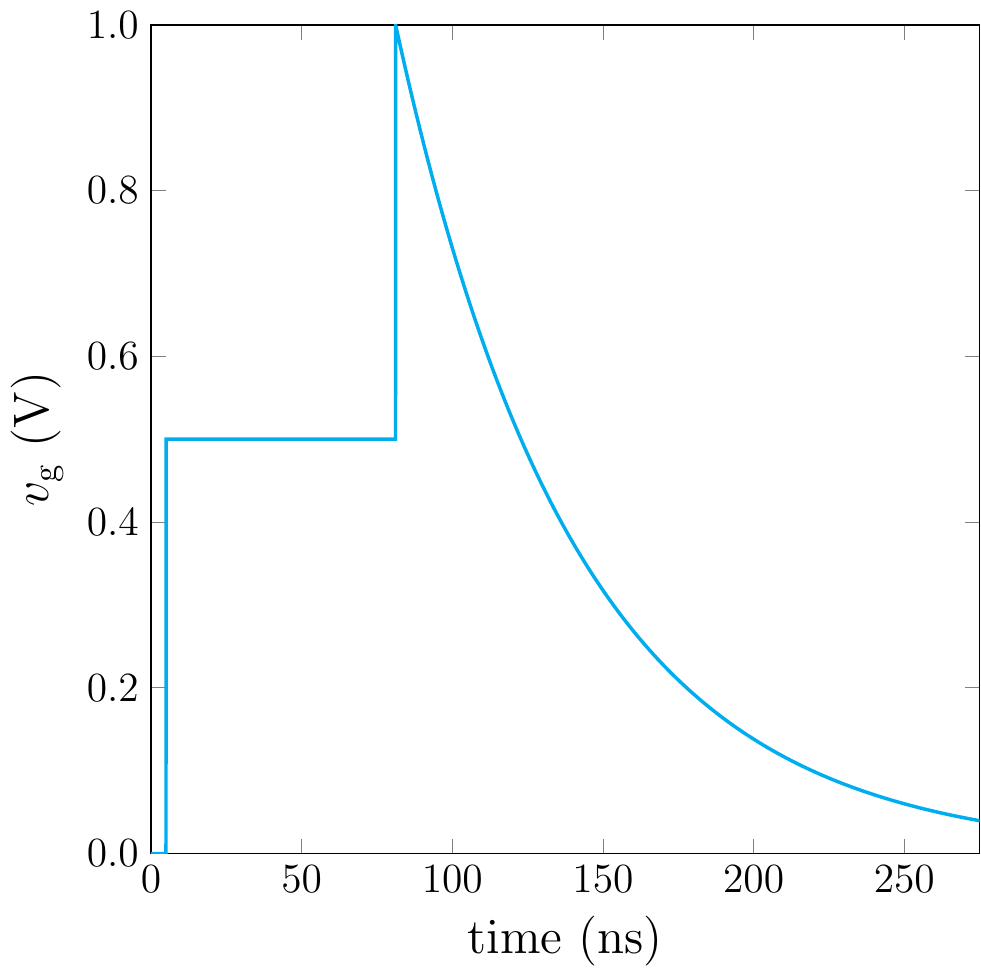} & ~\qquad~ & (b) \includegraphics[height=0.9\columnwidth]{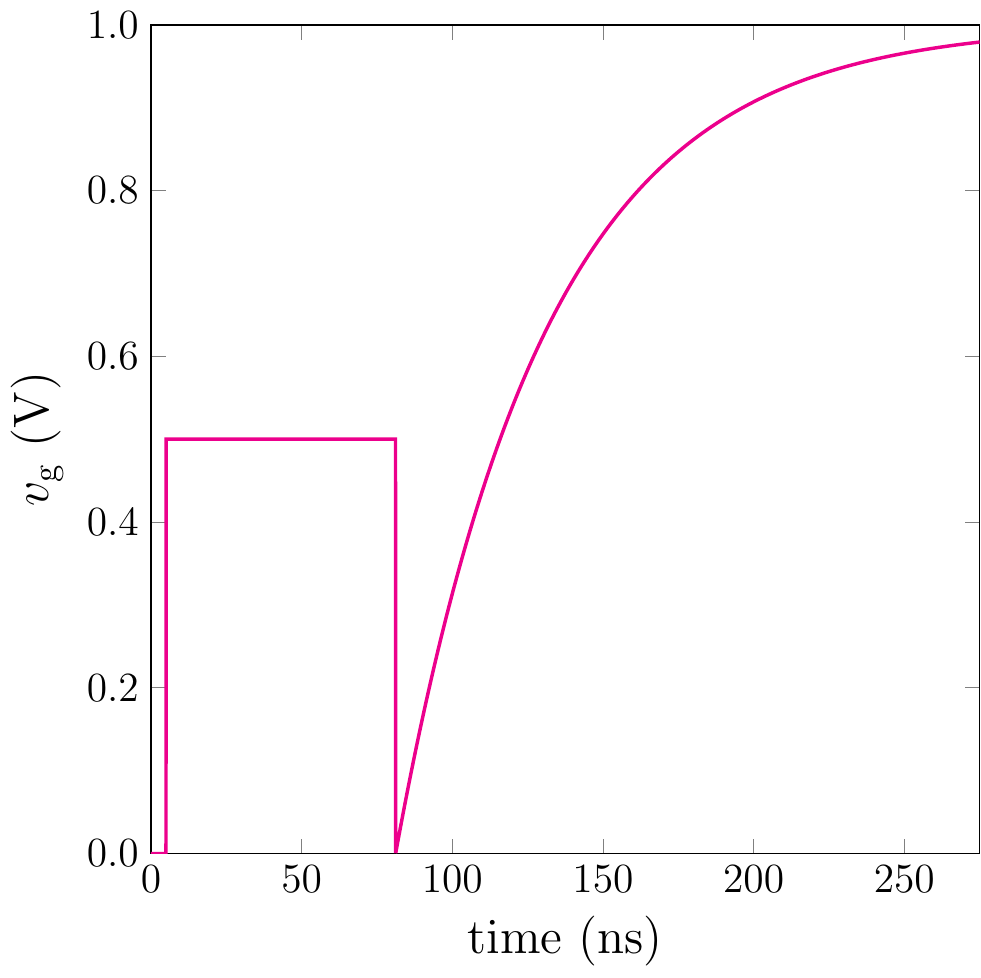}
\end{tabular}
\caption{Transient response of a transmission line to a voltage step with $Z_\mathrm{g}=50\rm\ \Omega$ and (a) an inductive and (b) a capacitive load termination.\label{fig:reactive}}
\end{figure*}

Taking the inverse Laplace transform of Eq.~(\ref{eq:ZL3}) gives the desired transient response.  Once again, we make use of the translation theorem given by Eq.~(\ref{eq:invTrans}) to evaluate the required transform:
\begin{multline}
v_\mathrm{g}(t) = \frac{V_0}{2}\Bigg\{u\left(t - t_\mathrm{c}\right) \Bigg.\\
\Bigg. + u\left[t - \left(t_\mathrm{c} + 2\ell/v_0\right)\right]\left[2e^{-\left[t - \left(t_\mathrm{c} + 2\ell/v_0\right)\right]/\tau_\mathrm{L}} - 1\right]\Bigg\}
\end{multline}
Note that the $\tau\to\infty$ and $\tau\to 0$ limits recover the expected open circuit and short circuit results given by Eq.~(\ref{eq:solnB}) for $\Gamma_\mathrm{L} = \pm 1$.  Figure~\ref{fig:reactive}(a) shows the transient response of a lossless transmission line terminated by an inductive load with $L = 3\rm\ \mu H$.  The other parameters used were $\ell = 8\rm\ m$, $v_0 = 0.7c$, $Z_\mathrm{g} = Z_\mathrm{c} = 50\rm\ \Omega$, $V_0 = 1\rm\ V$ and $t_\mathrm{c} = 5\rm\ ns$. 

A similar analysis for a capacitive load $Z_\mathrm{L} = \left(sC\right)^{-1}$ leads to:
\begin{multline}
v_\mathrm{g}(t) = \frac{V_0}{2}\Bigg\{u\left(t - t_\mathrm{c}\right) \Bigg.\\
\Bigg. + u\left[t - \left(t_\mathrm{c} + 2\ell/v_0\right)\right]\left[1 - 2e^{-\left[t - \left(t_\mathrm{c} + 2\ell/v_0\right)\right]/\tau_\mathrm{C}}\right]\Bigg\},
\end{multline}
where the capacitive time constant is given by $\tau_\mathrm{C} = Z_\mathrm{c}C$.  Figure~\ref{fig:reactive}(b) plots the transient response using the same parameters used for part (a) of the figure and $C = 1\rm\ nF$.

\section{Conclusions}\vspace{-5mm}
We have calculated the transient response of lossless transmission lines.  Despite their simplicity, the calculated and measured transient responses can be quite intricate and rich in physics.  In our approach, we first considered the transmission lines in the complex frequency domain which facilitated the use of basic circuit-analysis techniques.  Inverse Laplace transforms were then used to determine the desired time-domain response.  This approach assumes that one is able to (1) determine the $s$-domain representation of the signal generator's output and (2) analytically evaluate the required inverse Laplace transform.  If a closed-form solution to the inverse Laplace transform is not possible, then numerical approaches to approximately evaluate the required transform can be employed~\cite{Valsa:1998, Bobowski:2020}.  In future work, we hope to analytically solve for the transient response of low-loss transmission lines.


\bibliographystyle{apsrev} 
\bibliography{trans} 

\end{document}